\pdfoutput=1
\documentclass[onecolumn]{sdl}

\usepackage{multirow}

\newcommand{\beq}{\begin{equation}}
\newcommand{\eeq}{\end{equation}}
\newcommand{\beqn}{\begin{eqnarray}}
\newcommand{\eeqn}{\end{eqnarray}}

\def\0{\mbox{\tiny $0$}}
\def\1{\mbox{\tiny $1$}}
\def\2{\mbox{\tiny $2$}}
\def\3{\mbox{\tiny $3$}}
\def\4{\mbox{\tiny $4$}}
\def\5{\mbox{\tiny $5$}}
\def\6{\mbox{\tiny $6$}}
\def\7{\mbox{\tiny $7$}}
\def\8{\mbox{\tiny $8$}}
\def\9{\mbox{\tiny $9$}}

\logo{
\colorbox{DarkGoldenrod}{\color{white}$\mathbf{\Sigma\hspace*{0.06cm} \delta \hspace*{0.04cm} \Lambda}$}
}

\journal{\shadowtext{\textbf{\color{DarkRed} International Journal of Modern Physics E}} \, \textbf{27}, 1850039-18 (2018).}

\titlelines{3}
\title{Analytical and numerical\\ analysis of the complete\\ Lipkin-Meshkov-Glick Hamiltonian}

\imgbgabstract{The Lipkin-Meshkov-Glick is a simple, but not trivial,
model of a quantum many-body system which allows us to solve the many-body
Schr\"odinger equation without making any approximation. The
model, which in its
unperturbed case is composed only by two energy levels,
includes two interacting terms. A first one, the $V$ interaction, which promotes or
degrade pairs of particles, and a second one, the $W$ interaction, which scatters one
particle in the upper and another in the lower energy level.
In comparing this model with other approximation methods,
the $W$ term interaction is often set to zero. In this paper,
we show how the presence of this interaction changes
the global structure of the system, generates
degeneracies between the various eigenstates and modifies
the energy eigenvalues structure. We present analytical solutions for systems of
two and three particles and, for some specific cases,
also for four, six  and eight  particles. The solutions for
systems with more than eight particles are only numerical
but their behaviour can be well understood by considering
the extrapolations of the analytical results.
Of particular interest it is the study of how the $W$ interaction
affects the energy gap between the ground state and the first-excited  state.}

\author{
\names{Giampaolo Co$^{\prime\,1}$ and  Stefano De Leo$^{2,a}$}
\affiliation{$^{1}$Department of Mathematics and Physics, University of Salento, Italy}
\affiliation{$^{2}$Department of Applied Mathematics, State University of Campinas, Brazil}
\email{$^{a}$deleo@ime.unicamp.br}
}

\begin{document}

\sdlmaketitle

\section{Introduction}

The goal of the non relativistic many-body physics is to solve
a Schr\"odinger equation which describes a system composed
by many particles\cite{book1,book2}. Since two and, eventually, three-particles systems
are used to define the effective interaction between the particles,
in this context, ''many" means
more than three. The techniques to solve the many-body
Schr\"odinger equation without approximations are rather involved
and, usually, limited in the number of particles composing the systems.
For this reason, the methodologies most commonly used to solve
the quantum many-body problem are based on some simplifying
approximations.
One of the key problems of this branch of physics is the possibility
of testing the validity of these approximations.
With this aim, in the middle '60s of the last
century, Lipkin, Meshkov, and Glick (LMG) \cite{lip65a,lip65b,lip65c}
built a many-fermion system for which the Schr\"odinger equation can
be solved without approximations.
In this model, the non interacting, one-body, part of
the hamiltonian has only two energy levels, which are occupied by fermions.  The exact description of this system has been
compared with those obtained by using effective theories \cite{ap0,ap1,ap2,ap3,ap4}.
This model is a beautiful example of application of basic quantum
mechanics, and one of the few examples of many-body systems which can
be studied in relatively simple manner, without making approximations.

In this article, we consider the two level model containing in its
hamiltonian both the $V$ and the $W$ interaction terms.
After showing how to obtain the solutions of the one-body part
of the hamiltonian, we present a simple derivation of the exact solutions
for the complete LGM  hamiltonian. We show analytical expressions for system
of two and three particles
and numerical results for those with more particles. We discuss
the role of the various terms of the hamiltonian, and some general features
of the solutions, mainly those concerning the
degeneracies of some eigenstates.

\section{The Lipkin-Meshkov-Glick model}
\label{sec:LMG}

The original LMG model \cite{lip65a} considers
a hamiltonian of the type
\begin{equation}
H_{_{\mathrm{LMG}}} = E_{_0}\,J_{_0}  + \frac{V}{2}\,\left(\,\frac{J_{_{+}}^{^{\,2}} + J_{_-}^{^{\,2}}}{2}\,\right) +
\frac{W}{2}\, \left(\,\frac{J_{_+} J_{_-}\, + J_{_-} J_{_+}}{2}\,\right)\,\,.
\label{eq:LMGham}
\end{equation}
Nevertheless,
since the model was often used to test the treatment of
ground-state correlation in the random phase approximation,
the LMG  hamiltonian  is  usually
simplified by setting $W=0$.
Another possible choice for the $W$ interaction parameter
is given by $W=V$. In this case the hamiltonian can be
easily solved and its eigenvalues and eigenvectors are
used to study the instability of the Hartree-Fock state
against collective oscillations.

The aim of this paper is to present a detailed
discussion of the hamiltonian (\ref{eq:LMGham})
without any assumption or constraint on the $W$ interation term.

We start our study by
considering a system composed by $N$ non-interacting
particles. The configuration space available to each
particle is composed by only two states whose
energy values are separated by an amount $E_{_0}$,
\begin{equation}
\left\{\,-\,\frac{E_{_0}}{2}\,,\,\frac{E_{_0}}{2}\,\right\}\,\,.
\end{equation}
When the particles do not interact with each other
only the one-body part of the hamiltonian is present, $E_{_0}J_{_{0}}$.
In this case, the ground state of the system
is that where all the $N$ particles are lying on the lowest
state. The excited states are obtained by promoting
particles from the lower to the higher of the two energy levels.

A convenient representation of the many-body states
of this system is obtained by defining the
quantum numbers $j= N/2$ and  $m$, this latter one
can assume the values
\[
m=-j,\,-j+1\,,...\,,\,j-1,\,j\,\,.
\]
The eigenstates  of the unperturbed hamiltonian ($V=W=0$) then satisfy
\begin{equation}
J_{_0}\,|j,m\rangle = m\,|j,m\rangle\,\,.
\label{eq:kzero}
\end{equation}
The global number of possible eigenstates of this system is $2j+1$,
and the energies of these eigenstates are $m \,E_{_0}$.
A pictorial representation of the eigenstates of systems with
$N=2$, $3$, and $4$ particles is given in Table \ref{table:states}.

The  $J_{_+}$ operator removes a particle from the lower level and
put it on the higher one. The lowering operator, $J_{_-}$,
acts in the opposite
direction. The action of these operators on the eigenstates of
$J_{_0}$, Eq. (\ref{eq:kzero}), is
\begin{equation}
J_{_\pm} |j,m\rangle  =   \sqrt{j(j+1)-m(m\pm 1)}\,\,|j,m\pm1\rangle\,\,.
\label{eq:kpm}
\end{equation}

The $J^{^{\,2}}_{_+}$ term promotes particle pairs
from the lower to the upper state, while the $J^{^{\,2}}_{_-}$ term
operates in the opposite direction.
The $J_{_+}J_{_-}$ operator and its hermitian conjugate promotes
one particle and lowers another one.
The constants $V$ and $W$ are the strengths of the interactions between
the two particles involved in the processes described above.

We prefer to work with dimensionless quantities and with attractive interactions
\[\left(\,V\,,\,W\,\right)=-\,\left(\,|V|\,,\,|W|\,\right)\,\,,\]
therefore we divide the LMG hamiltonian (\ref{eq:LMGham}) by a constant factor
related to the the two-level model energy eigenvalue
\begin{equation}
{\mathcal H}\,=\,\frac{H_{_{\mathrm{LMG}}}}{E_{_0}} \,=\,
J_{_0}  - \frac{\,\nu}{2}\,\left(J_{_+}^{^{\,2}} + J_{_-}^{^{\,2}}   \right) -
\frac{\,\omega}{2}\, \left(J_{_+} J_{_-}\, + J_{_-} J_{_+}\, \right),
\label{eq:hamnew}
\end{equation}
where we have defined
\begin{equation*}
\nu        = |V|\, / \, E_{_0}  \,\,\,\,\,\,\,\, {\mathrm{and}} \,\,\,\,\,\,\,\,
\omega = |W|\, / \, E_{_0}
\,\,\,.
\end{equation*}
Many of the of matrix elements of ${\mathcal H}$  between the eigenstates
of $J_{_0}$ are zero, those different from zero are
\begin{eqnarray}
\langle j, m | {\mathcal H}| j, m \rangle &=&   m - \omega [\,j(j+1) - m^{2}\,], \nonumber \\
 \langle j, m |  {\mathcal H}| j, m +2 \rangle & =&
   -\,\frac{\nu}{2}\,\sqrt{[\,j\,(j+1) - (m+1)\,m\,]\,\,[\,j\,(j+1)-(m+2)\,(m+1)\,]},\nonumber \\
 \langle j, m + 2 |  {\mathcal H}_{_{\mathrm{LMG}}}| j, m \rangle &=&
 \langle j, m |  {\mathcal H} | j, m +2 \rangle .
\end{eqnarray}
We define a new quantum number $r$, related to $N$ and $m$
by the expression
\begin{equation}
m=\frac{N}{2}-r+1\,\,\,\,\,(r=1,2,...,N+1),
\end{equation}
and we express the above matrix elements as a function of the particle
number $N$ and of the interaction parameters $\nu$ and $\omega$,
\begin{equation}
{\mathcal H}_{rs}^{^{[N]}}=\left\{\begin{array}{ccl}
\displaystyle{\frac{N}{2}}-r+1 - \left(N\,r-\frac{N}{2}-r^{2}+2\,r-1\right)\,\omega &\,\,\,\,\,\,\,& \mbox{for}\,\,s=r\,\,;\,\,(r=1,2,...,N+1),\\
 & & \\
-\,\displaystyle{\frac{\sqrt{(N-r)(N-r+1)(r+1)\,r}}{2}}\,\,\nu & &\mbox{for}\,\,  s=r+2
\,\,;\,\,(r=1,2,...,N-1),\\
 & & \\
-\,\displaystyle{\frac{\sqrt{(N-s)(N-s+1)(s+1)\,s}}{2}}\,\,\nu  & &\mbox{for}\,\,  r=s+2
\,\,;\,\,(s=1,2,...,N-1),\\
 & & \\
0 & & \mbox{otherwise}.
\end{array}\right.
\label{eq:matrix}
\end{equation}
The solution of the Schr\"odinger equation consists in diagonalizing the above matrix.
In this manner, we obtain the eigenvalues of the hamiltonian
(\ref{eq:hamnew}) in terms of the $N$, $r$ quantum numbers and of the $\nu$ and
$\omega$ stregths. The corresponding eigenvectors are expressed in the basis
formed by the eigenstates of $J_{_0}$.

\section{Solutions for $\boldsymbol{N=2}$ and $\boldsymbol{N=3}$.}
\label{sec:lipkinA}

For the systems composed by 2 or 3 particles we found
analytical expressions of eiegenvalues and eigenvectors
of the full hamiltonian (\ref{eq:hamnew}).
We obtained analytical expressions of eiegenvalues and eigenvectors
also for systems composed by 4, 6 and 8 particles, but only
when in the hamiltonian we set $\omega=0$.

We exploit the
analytical form of solutions
to investigate
the effects of the terms proportional to $\nu$ and $\omega$
as they evolve with
the increasing number of particles
and to suggest how to analyse the numerical solutions.

The explicit expression of the hamiltonian matrix for the system
composed by two particles is
\begin{equation}
{\mathcal H}^{^{[2]}}=\left(\begin{array}{ccc} 1-\omega & 0 & -\nu  \\ 0 & -\,2\,\omega & 0 \\ -\,\nu & 0 & -\,(1 +\omega)  \end{array} \right)
\,\,,
\end{equation}
where we have indicated in the upperindex between square brackets
the number of particles composing the system. We write the
eigenvalues, ${\mathcal E}^{^{[2]}}(\nu,\omega)$
and the eigenvectors, ${\mathcal V}^{^{[2]}}(\nu,\omega)$ of this system as
\begin{equation}
\left\{\,{\mathcal E}^{^{[2]}}(\nu,\omega)\,;\,{\mathcal V}^{^{[2]}}(\nu,\omega)\,\right\}=\left\{
\begin{array}{lcl}
+\, \sqrt{1+  \nu^{^2}} - \omega &\,\,\,;\,\,\,&
\left(\,\displaystyle{\frac{1+ \sqrt{1+  \nu^{^2}}}{2}} \,,\,0\,,\,-\,\frac{\nu}{2}\,\right)^\top\\
 -\,2\,\omega &;&    \left(\,0 \,,\,1\,,\,0\,\right)^t \\
 -\,\sqrt{1+  \nu^{^{2}}} - \omega &;&
 \left(\,\displaystyle{\frac{\sqrt{1+  \nu^{^2}}-1}{\nu}} \,,\,0\,,\,1\,\right)^\top \end{array} \right\},
\end{equation}
where $\top$ indicates the  transpose which, in this case,
transform the rows into columns.
We have selected a global normalisation of the eigenvectors
which allows us to recover, in the limit where both $\nu$ and $\omega$ are
zero, the simple form $(1,0,0)^\top$, $(0,1,0)^\top$ and $(0,0,1)^\top$.

For the case of three particles, we found the following
expression of the hamiltonian  matrix
\begin{equation}
H^{^{[3]}}=\left(\begin{array}{cccc}
\frac{3}{2}\,(1-\omega) & 0 & -\,\sqrt{3}\,\,\nu & 0  \\
0 &  \frac{1}{2}\,(1-7\,\omega) & 0 & -\,\,\sqrt{3}\,\,\nu \\
 -\,\,\sqrt{3}\,\,\nu & 0 &  -\,\frac{1}{2}\,(1+7\,\omega)  & 0 \\
 0 & -\,\,\sqrt{3}\,\,\nu & 0 & -\,\frac{3}{2}\,(1+\omega)
\end{array} \right).
\end{equation}
The energy eigenvalues are
\begin{equation}
{\mathcal E}^{^{[3]}}(\nu,\omega) = \left[
\begin{array}{c}
{\mathcal E}^{^{[3]}}_{_{4}}(\nu,\omega)\\
{\mathcal E}^{^{[3]}}_{_{3}}(\nu,\omega)\\
{\mathcal E}^{^{[3]}}_{_{2}}(\nu,\omega)\\
{\mathcal E}^{^{[3]}}_{_{1}}(\nu,\omega)
\end{array}
   \right] = \,\frac{1}{2}  \, \left[\begin{array}{l}
 +1 + 2\,\sqrt{(1+\omega)^{^2} + 3\,\nu^{^{2}}} - 5\,\omega\\
 -1 + 2\,\sqrt{(1-\omega)^{^2}+  3\,\nu^{^{2}}}- 5\,\omega\\
  +1 - 2\,\sqrt{(1+\omega)^{^2} + 3\,\nu^{^{2}}}- 5\,\omega\\
   -1 - 2\,\sqrt{(1-\omega)^{^2}+  3\,\nu^{^{2}}}- 5\,\omega
\end{array}\right]\,\,,
\label{eq:ene3p}
\end{equation}
where they are ordered with increasing values for $\nu=0$ and
$\omega=0$.

We show in Fig. \ref{fig:fig1} the evolution of these four eigenvalues as a function
of $\nu$ for 6 selected values of $\omega$.
Since we are interest in studying when degeneracy appear,
we write the energy differences between the
$r$-th and $s$-th levels as
\[\Delta_{_{r,s}}^{^{[N]}}(\nu,\omega)=  {\mathcal E}^{^{[N]}}_{_{r}}(\nu,\omega)- {\mathcal E}^{^{[N]}}_{_{s}}(\nu,\omega)\,\,. \]
In the case of $\omega=0$ we recover the traditional LMG
model \cite{lip65a,lip65b,lip65c}. For this case we have
\begin{equation}
\Delta^{^{[3]}}(\nu,0)=\left[  \begin{array}{l}
\Delta_{_{4,3}}^{^{[3]}}(\nu,0)\\
\Delta_{_{3,2}}^{^{[3]}}(\nu,0)\\
\Delta_{_{2,1}}^{^{[3]}}(\nu,0)
\end{array}  \right] =
\left[  \begin{array}{c} 1\\ -\,1 + 2\,\sqrt{1 + 3\,\nu^{^{2}}}  \\ 1  \end{array} \right]\,\,.
\label{eq:e3diff}
\end{equation}
We observe that the energy gap between the first/second and third/fourth
energy levels remains unchanged by increasing the value of $\nu$. The energy gap
between the second/third energy level increases when we increase the
$\nu$ value  and, consequently, the energy levels in the traditional
LMG model never cross with each other, see the panel (a) of Fig.~\ref{fig:fig1}.

Values of $\omega$ different from zero break this symmetry.
We show in Fig.~\ref{fig:fig2} the evolution of the four egeinenergies
as a function of $\omega$ for 6 values of $\nu$.
When $\nu=0$,
see the panel (a) of Fig.~\ref{fig:fig2}, we find

\begin{equation}
\Delta^{^{[3]}}(0,\omega) =
\left[
\begin{array}{l}
\Delta_{_{4,3}}^{^{[3]}}(0,\omega) \\
\Delta_{_{3,2}}^{^{[3]}}(0,\omega)\\
\Delta_{_{2,1}}^{^{[3]}}(0,\omega)
\end{array}
\right] =
\left[  \begin{array}{c} 1 + 2 \,\omega\\ 1  \\ 1 -2\,\omega \end{array} \right].
\end{equation}
In this case, the values of the first and second energy eigenvalues cross
at $\omega=1/2$. In general, the crossing point, for a given value of $\nu$,
satisfies the equation
\beq
\Delta_{_{2,1}}^{^{[3]}}(\nu,\omega) =
{\mathcal E}_{_{2}}^{^{[3]}}(\nu,\omega) -
{\mathcal E}_{_{1}}^{^{[3]}}(\nu,\omega) =0,
\label{eq:delta3}
\eeq
which implies that the $\omega$ and $\nu$ terms are related by the expression
\begin{equation}
\omega^{^{[3]}}_{_{2,1}} = \sqrt{\nu^{^{2}}+\,\,\frac{1}{4}\,}\,\,.
\label{eq:nudelta3}
\end{equation}
%
The eigenvectors of the 3 particle system related to the four
eigenvalues can be expressed as
\begin{equation}
{\mathcal V}^{^{[3]}}(\nu,\omega) ={\cal A}(\nu,\omega) \,
\left\{\begin{array}{l}
\left(\, \displaystyle{\frac{1 + \sqrt{1+a_+^2(\nu,\omega)}}{2}} \,,\,0\,,\,-\,\frac{a_+(\nu,\omega)}{2}\,,\,0\,\right)^\top,\\
\left(\,0\,,\, \displaystyle{\frac{1 + \sqrt{1+a_-^2(\nu,\omega)}}{2}} \,,\,0\,,\,-\,\frac{a_-(\nu,\omega)}{2}\,\right)^\top,\\
\left(\, \displaystyle{\frac{\sqrt{1+a_+^2(\nu,\omega)}-1}{a_+(\nu,\omega)}} \,,\,0\,,\,1\,,\,0\,\right)^\top,\\
\left(\, 0\,,\,\displaystyle{\frac{\sqrt{1+a_-^2(\nu,\omega)}-1}{a_-(\nu,\omega)}} \,,\,0\,,\,1\,\right)^\top.
\end{array}\right.
\label{eq:estate3}
\end{equation}
where we have defined the quantity
\beq
a_{\pm}(\nu,\mu)=\frac{\sqrt{3}\,\nu}{1\pm \omega}.
\eeq
In the limit of $\nu=\omega=0$ the eigenvectors assume the
simple form $(0,0,0,1)^\top$, etc.  and the normalisation constant
\beq
{\cal A}(\nu,\omega)
= \displaystyle{\frac{a_-(\nu,\omega)}{\sqrt{2\,\sqrt{1+a_-^2(\nu,\omega)}\,\left(\,\sqrt{1+a_-^2(\nu,\omega)}-1\,\right)}}}
\eeq
is chosen to guarantee that each eigenvector is normalised to one.

We studied how the eigenvectors changes as a function of the interaction
strengths $\nu$ and $\omega$. As example of these changes, we show
in Fig.~\ref{fig:fig3}, for the lowest energy state,
the behaviour of the squares of the two non zero
components. We remark that for $\nu\ll|1-\omega|$, i.e.
in the limit $a_-^2(\nu,\omega) \rightarrow 0$,  the lowest energy
eigenvector  assumes the values
\[(\,0\,,\,0\,,\,0\,,\,1\,)^\top.\]
For $\nu=|1-\omega|$, i.e. when $a_-^2(\nu,\omega) \rightarrow 3$, we find
\[\left(\,0\,,\,0.25\,,\,0\,,\,0.75\,\right)^\top.\]
Finally, in the limit for $a_-^2(\nu,\omega) \rightarrow \infty$ ($\nu\gg|1-\omega|$),
we obtain for the energy eigenvector the representation
\[\left(\,0\,,\,0.5\,,\,0\,,\,0.5\,\right)^\top.\]
In all the panels of Fig. \ref{fig:fig3} and  \ref{fig:fig4} the full black line
and the red dashed line represent, respectively,
the second and fourth component of the lowest energy eigenvector.
The difference between the two components start from 1 and tends to zero
when the value of $\nu$ inceases.
This can be see in all the panels of Fig. \ref{fig:fig3} and \ref{fig:fig4} .
The interaction depending on $\omega$ potential accelerate this behaviour
when the values of $\omega$ are between 0 and 1, see Fig. \ref{fig:fig3}.
At $\omega=1$, the trend is inverted, as it is shown in Fig. \ref{fig:fig4},
and the difference between these two components of the lowest
energy eigenstate tends to 1. These two components are equal,
and assume the value of 0.5, for $\omega=1\mp\nu$, see Fig. \ref{fig:fig4}.

\section{Solutions for $\boldsymbol{N \ge 4}$ }
\label{sec:lipkinB}

We obtain analytical expressions for the solutions of the systems
with  $N=4,6,8$ when we set $\omega=0$,
\begin{eqnarray}
{\mathcal E}^{^{[4]}}(\nu,0)  & = & 0,\quad \pm\,\sqrt{1+9\, \nu^{^{2}}},\quad \pm\,2\, \sqrt{1+3\, \nu^{^{2}}},\nonumber \\
{\mathcal E}^{^{[6]}}(\nu,0)& = & 0,\quad \pm\,2\, \sqrt{1+15 \,\nu^{^{2}}},\quad \pm\,\sqrt{5+33\, \nu^{^{2}}\pm 4 \sqrt{1+6\,\nu^{^{2}}+54\, \nu^{^{4}}}},\\
{\mathcal E}^{^{[8]}}(\nu,0)& = &   0,\quad \pm\,\sqrt{5+113\,\nu^{^{2}}\pm 4 \sqrt{1+38\,\nu^{^{2}}+550\,\nu^{^{4}}}},
\quad \pm\,\sqrt{10+118\,\nu^{^{2}}\pm 6 \sqrt{1-2\,\nu^{^{2}}+225\,\nu^{^{4}}}}.\nonumber
\end{eqnarray}
The above expressions correct those given in the original paper of
Lipkin et al.~\cite{lip65a} where a
factor $4$ in $E^{^{[6]}}$, and  a factor
$6$ in $E^{^{[8]}}$, are missing.

The energy differences between the first excited state and the ground state
in each of these systems are
\begin{eqnarray*}
\Delta_{_{2,1}}^{^{[4]}}(\nu,0) & = &  2\, \sqrt{1+3\, \nu^{^{2}}} - \sqrt{1+9\, \nu^{^{2}}},\\
\Delta_{_{2,1}}^{^{[6]}}(\nu,0) & = &   \sqrt{5+33\, \nu^{^{2}}+ 4 \sqrt{1+6\,\nu^{^{2}}
+54\, \nu^{^{4}}}}
-\,2\, \sqrt{1+15 \,\nu^{^{2}}},\\
\Delta_{_{2,1}}^{^{[8]}}(\nu,0) & = &  \sqrt{10+118\,\nu^{^{2}}+ 6 \sqrt{1-2\,\nu^{^{2}}
+225\,\nu^{^{4}}}} - \sqrt{5+113\,\nu^{^{2}}+ 4 \sqrt{1+38\,\nu^{^{2}}+550\,\nu^{^{4}}}}.
\end{eqnarray*}
As expected, for a free hamiltonian we have $\Delta_{_{2,1}}^{^{[4]}}(0,0)=\Delta_{_{2,1}}^{^{[6]}}(0,0)=\Delta_{_{2,1}}^{^{[8]}}(0,0)=1$.

The above results indicate that, for a given $\nu$ value,
these quantities become smaller with the increase of the
particle number $N$. For example, in the case $\nu=1$,
we obtain
\begin{eqnarray*}
\Delta_{_{2,1}}^{^{[4]}}(1,0) & = &  4 - \sqrt{10}\,\,\approx\,\, 0.838\,\,,\\
\Delta_{_{2,1}}^{^{[6]}}(1,0) & = &   \sqrt{38+ 4 \sqrt{61}}
-\,8\,\,\approx\,\, 0.321\,\,,\\
\Delta_{_{2,1}}^{^{[8]}}(1,0) & = &  2\,\sqrt{43 + 6 \sqrt{14}} - \sqrt{118 + 4 \sqrt{589}}
\,\,\approx\,\,0.093
\,\,\,.
\end{eqnarray*}

The energy eigenvalues for the cases investigated are shown in Fig. \ref{fig:fig5}
for $0 \le \nu \le 1$ when $\omega=0$. The various lines never cross
with each other and show a symmetry with respect to the zero value. The differences
from this value increase with the increase of the value of $\nu$.
In the figure it is evident the lowering of the $\Delta^{^{[N]}}_{_{2,1}}(\nu,0)$ values
with increasing of $N$ and of $\nu$.

We investigate the role of $\omega$ in the hamiltonian (\ref{eq:LMGham})
by considering the case of $\nu=0$.
We found that, for a system composed by $N$ particles, the eigenvalues
of the $m$ state can be expressed as
\beq
\mathcal{E}^{^{[N]}}_{m} (0,\omega)
=   m - \omega \,[\,j\,(j+1) - m^{^2}\,]
\,\,\,,
\label{eq:engen}
\eeq
where $m$ can assume the integer values $(-j,-j+1,...,j-1,j)$
with $j=N/2$.

The energy difference between the level $m+p$ and the level $m$ is
\begin{equation}
\Delta^{^{[N]}}_{_{m+p,m}} = p\,[\,1\,+\,\omega\,(\,2\,m\,+\,p\,)\,]
\end{equation}
which is a constant, i.e. independent of  $\omega$, when $p=-2\,m$.

The energy level $r$ ($m=-N/2+r-1$) intercepts the  level $s$ ($p=s-r$)
when  $\omega$ acquires the value
\begin{equation}
\omega_{_{r,s}}^{^{[N]}}=\frac{1}{N-r-s+2}\,\,.
\label{eq:ors}
\end{equation}
We show in Fig.~\ref{fig:fig6} the behaviour of the energy eigenvalues
for the $N=4,6,8$ systems as a function of $\omega$, when $\nu=0$.
The vertical lines highlight the position of the crossing points
obtained in (\ref{eq:ors}).
This expression implies degeneracies for equal values of $r+s$.
For example, for  $N=8$, see Fig.~\ref{fig:fig6}(c), we have a triple
degeneracy when
\[\omega_{_{1,6}}^{^{[8]}}=\omega_{_{2,5}}^{^{[8]}}=\omega_{_{3,4}}^{^{[8]}}=\frac{1}{3}
\,\,\,\,\,\,\,{\mathrm{and}}\,\,\,\,\,\,\,
\omega_{_{1,7}}^{^{[8]}}=\omega_{_{2,6}}^{^{[8]}}=\omega_{_{3,5}}^{^{[8]}}=\frac{1}{2}\,\,,
\]
and a fourfold  degeneracy for
\[\omega_{_{18}}^{^{[8]}}=\omega_{_{27}}^{^{[8]}}=\omega_{_{36}}^{^{[8]}}=\omega_{_{45}}^{^{[8]}}=1\,\,.\]

We have numerically studied the solution for $N > 8$.
We show in Fig.~\ref{fig:fig7} the energy difference between the
two lowest states of a system composed by 10 interacting particles
as a function of the renormalised interaction strength $\nu$.
The various lines indicate the results obtained with different values
of $\omega$. We observe that by increasing the $\omega$ values,
i.e. the strength of the last term of the hamiltonian
(\ref{eq:LMGham}), this difference decreases, see panel (a) of
Fig.~\ref{fig:fig7}.

We found a relation between the $\nu$ and $\omega$ strengths
indicating when the two levels are degenerated. We generalise
the results obtained for the 3 particle system, Eq.~(\ref{eq:nudelta3}),
and we found the following relation for $N$ particles:
\begin{equation}
\omega^{^{[N]}}_{_{2,1}} = \sqrt{\nu^{^{2}}+\,\,\frac{1}{(N-1)^{^{2}}}\,}\,\,.
\label{eq:box}
\end{equation}

In the specific case presented in Fig.~\ref{fig:fig7}, $N=10$, we have
that for a strength $\omega=1/9$ the two levels become degenerated for
$\nu=0$, see the red line in the panel (b) of Fig.~\ref{fig:fig7}. For
stronger $\omega$ strengths the degeneracy appear at positive values
of $\omega$ as it is shown in the panel (b) of Fig.~\ref{fig:fig7} by
the black line.


In Fig.~\ref{fig:fig8} we compare some result of the $N=10$
system with those obtained for $N=20$ and $N=50$. The blue lines
indicate the $\Delta_{_{2,1}}^{^{[N]}}$ values when $\omega=0$, this is the
result of the
usual LMG model. We observe that the degeneracy of the two levels
appear at smaller values of the interaction strength $\omega$.

The red lines show the results obtained by including also the second
interaction term of the Hamiltonian. We have chosen the values
$\omega=1/(N-1)$
which satisfy Eq.~(\ref{eq:box}) when $\nu=0$.  We observe that
the degeneracies for $\nu > 0$ appear for values always smaller with
the increase of $N$.  It is worth to remark in the figure, the
differences in the $\nu$ scale between the different panels and the
amplification factors of the red lines.

\section{Conclusions}
In this article, we have investigated the two-level LMG model when the
hamiltonian includes a two-body interaction term which allows particle-hole
type of excitations. We have studied the effects of the interaction terms of
the LMG hamiltonian (\ref{eq:LMGham}) by changing $\omega$ and $\nu$,
the strengths of the two interaction terms.  The $\omega=0$ case reproduce
the traditional LMG model \cite{lip65a,lip65b,lip65c}.

For $N=2,3$ we obtain fully analytical solutions for both eigenvalues and
eigenvectors. The presence of the new interaction term implies the
degeneracy of some of the solutions.
By changing the values of the strengths $\nu$ and $\omega$
the energies of the lowest states is modified and the higher states
becomes a new ground state.

This feature is emphasised when the number of the particle composing the
system increases. We have pointed out cases when the energy differences
become always smaller with increasing the $\nu$ values but they are never
zero. In other cases, instead, it is evident the crossing between the different
eigenvalues resulting in a change of the ground state eigenvector.

We give in Eq. (\ref{eq:box}) a semi-empirical expression which relates
the zeros of the energy differences between the two lowest states of
a system with $N$ particles with the strength $\nu$ of the interaction.

\newpage

\begin{table}
{\large
\begin{center}
\begin{tabular}{|c||c|l|}
\hline
 \multirow{9}{*}{\color{blue}\,\,\textbf{N=2}\,\,}
 & & \\
 & \,\,{\color{green} $\bullet$}\,\,{\color{green} $\bullet$}\,\, & \color{blue} \normalsize $|\,1\,,\,1\,\rangle$ \\
 & & \\ \cline{2-3} & & \\ & {\color{green} $\bullet$}\,\,{\color{red} $\bullet$} & \color{blue} \normalsize $|\,1\,,\,0\,\rangle$ \\
 & & \\ \cline{2-3} & & \\  & {\color{red} $\bullet$}\,\,{\color{red} $\bullet$} & \color{blue} \normalsize $|\,1\,,\,\mbox{-}\,1\,\rangle$ \\ & & \\ \hline \end{tabular}
\begin{tabular}{|c||c|l|}
\hline
 \multirow{12}{*}{\color{blue}\,\,\textbf{N=3}\,\,} & & \\
& \,\,{\color{green} $\bullet$}\,\,{\color{green} $\bullet$}\,\,{\color{green} $\bullet$}\,\, & \color{blue} $|\,\frac{3}{2}\,,\,\frac{3}{2}\,\rangle$ \\ & & \\ \cline{2-3} & & \\
 & {\color{green} $\bullet$}\,\,{\color{green} $\bullet$}\,\,{\color{red} $\bullet$} & \color{blue} $|\,\frac{3}{2}\,,\,\frac{1}{2}\,\rangle$ \\  & & \\ \cline{2-3} & & \\
 & {\color{green} $\bullet$}\,\,{\color{red} $\bullet$}\,\,{\color{red} $\bullet$} & \color{blue} $|\,\frac{3}{2}\,,\,\mbox{-}\,\frac{1}{2}\,\rangle$ \\ & & \\  \cline{2-3} & & \\
 & {\color{red} $\bullet$}\,\,{\color{red} $\bullet$}\,\,{\color{red} $\bullet$} & \color{blue} $|\,\frac{3}{2}\,,\,\mbox{-}\,\frac{3}{2}\,\rangle$ \\ & & \\ \hline
\end{tabular}
\begin{tabular}{|c||c|l|}
\hline
 \multirow{15}{*}{\color{blue}\,\,\textbf{N=4}\,\,} & & \\
 & \,\,{\color{green} $\bullet$}\,\,{\color{green} $\bullet$}\,\,{\color{green} $\bullet$}\,\,{\color{green} $\bullet$}\,\, & \color{blue} \normalsize $|\,2\,,\,2\,\rangle$ \\ & & \\ \cline{2-3} & & \\
  & {\color{green} $\bullet$}\,\,{\color{green} $\bullet$}\,\,{\color{green} $\bullet$}\,\,{\color{red} $\bullet$} & \color{blue} \normalsize $|\,2\,,\,1\,\rangle$ \\ & & \\ \cline{2-3} & & \\
   & {\color{green} $\bullet$}\,\,{\color{green} $\bullet$}\,\,{\color{red} $\bullet$}\,\,{\color{red} $\bullet$} & \color{blue} \normalsize $|\,2\,,\,0\,\rangle$ \\ & & \\  \cline{2-3} & & \\
    & {\color{green} $\bullet$}\,\,{\color{red} $\bullet$}\,\,{\color{red} $\bullet$}\,\,{\color{red} $\bullet$} & \color{blue} \normalsize $|\,2\,,\,\mbox{-}\,1\,\rangle$ \\ & & \\ \cline{2-3} & & \\
     & {\color{red} $\bullet$}\,\,{\color{red} $\bullet$}\,\,{\color{red} $\bullet$}\,\,{\color{red} $\bullet$} & \color{blue} \normalsize $|\,2\,,\,\mbox{-}\,2\,\rangle$ \\ & & \\ \hline
\end{tabular}
\end{center}
}
\caption{Representation of the
eigenstates of the one-body hamiltonian
for systems  with $N=2$, $3$, and $4$ particles.
The number of green and red dots represents
the number of particles lying on the upper and on the lower
energy level, respectively.
}
\label{table:states}
\end{table}
%
%
%

\ColumnFigure{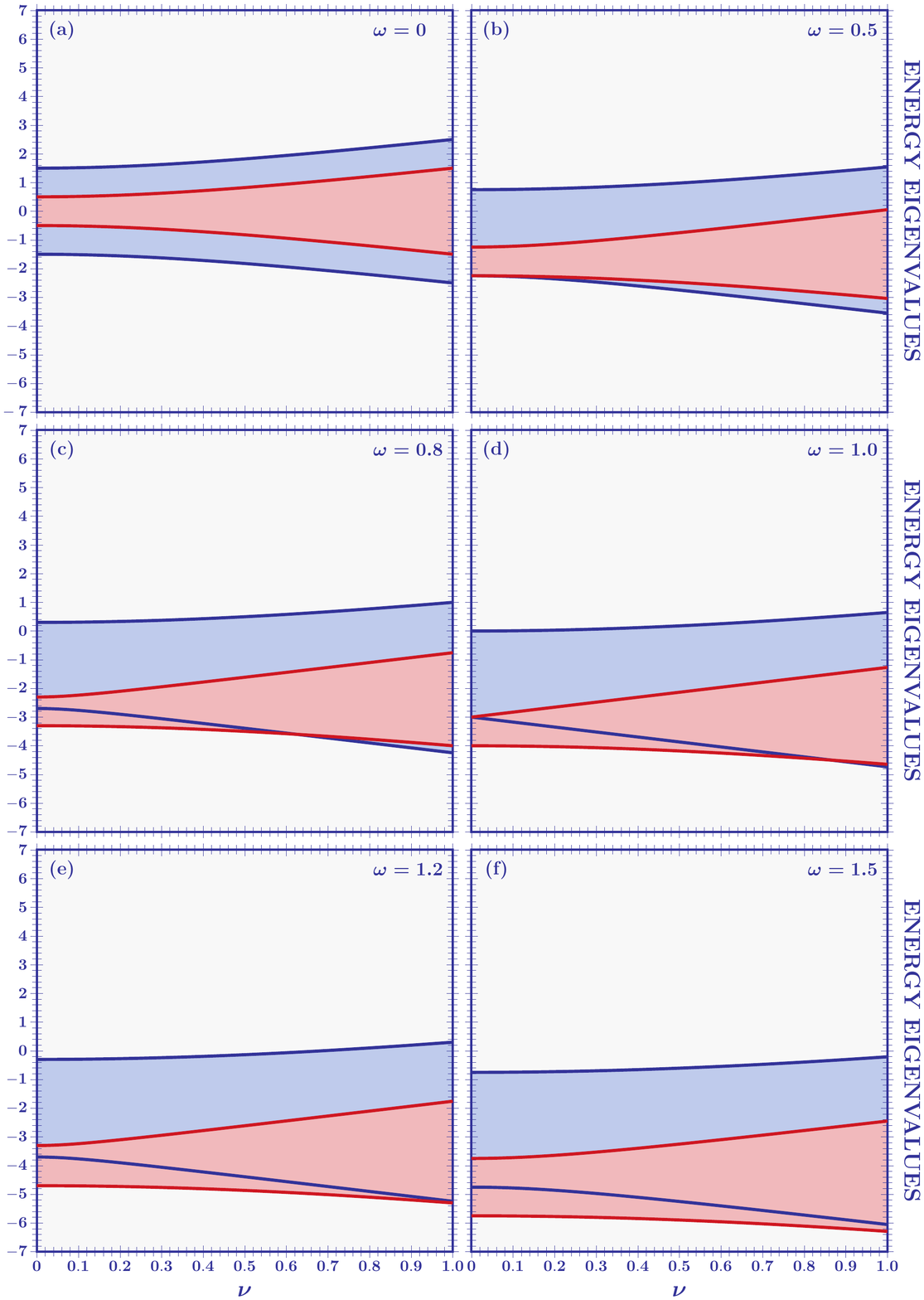}{Eigenenergies of a system of 3 particles, Eq.~(\ref{eq:ene3p}), as a function
of the strength $\nu$ for some fixed values of the strength $\omega$.
The blues and red lines identify the the specific eigenvalues.
The coloured areas are drawn to highlight the differences  between the eigenvalues.\label{fig:fig1}}

\ColumnFigure{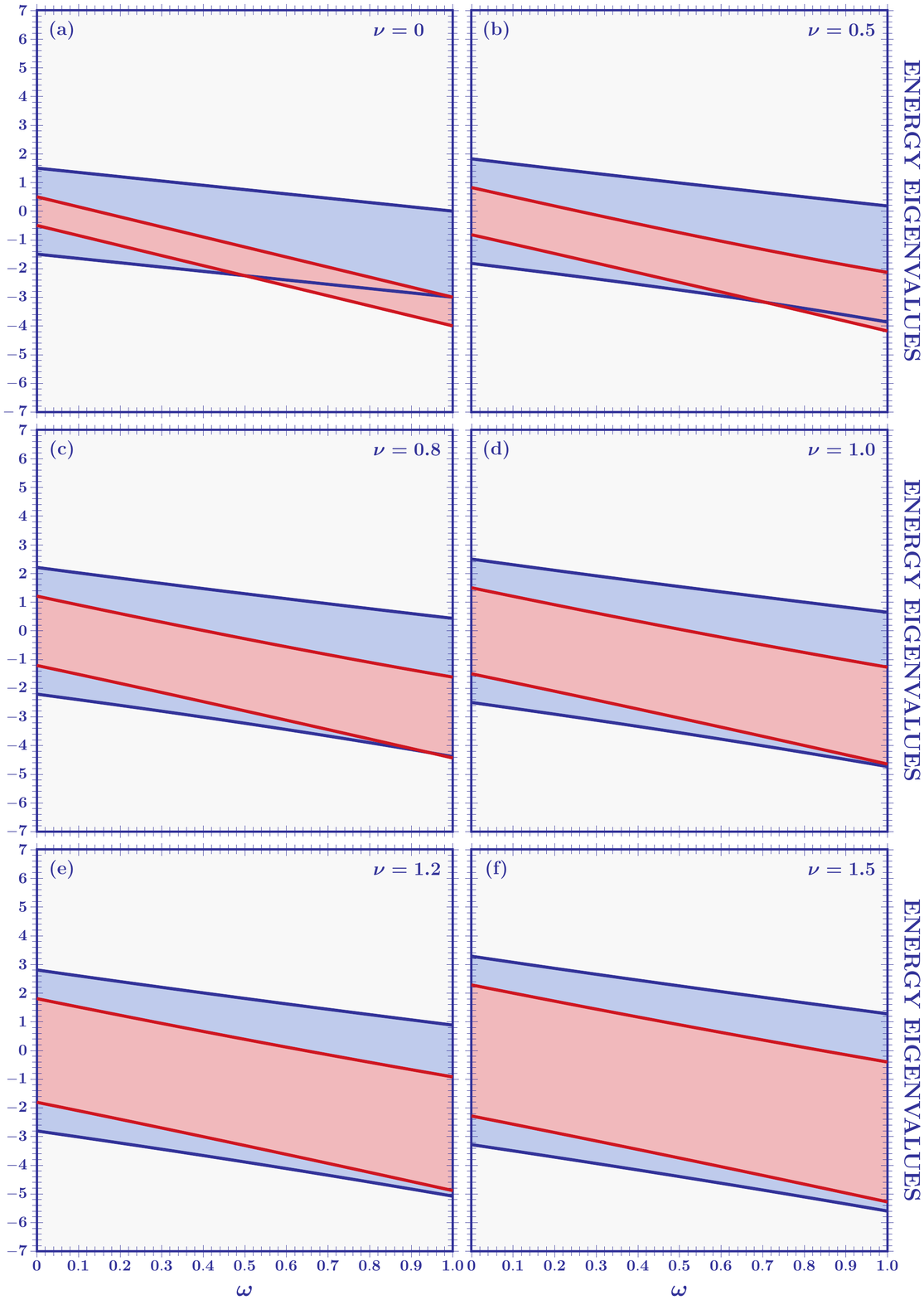}{The same as Fig. \ref{fig:fig1}, but as a function of the strength
$\omega$ for six values of $\nu$.\label{fig:fig2}}

\ColumnFigure{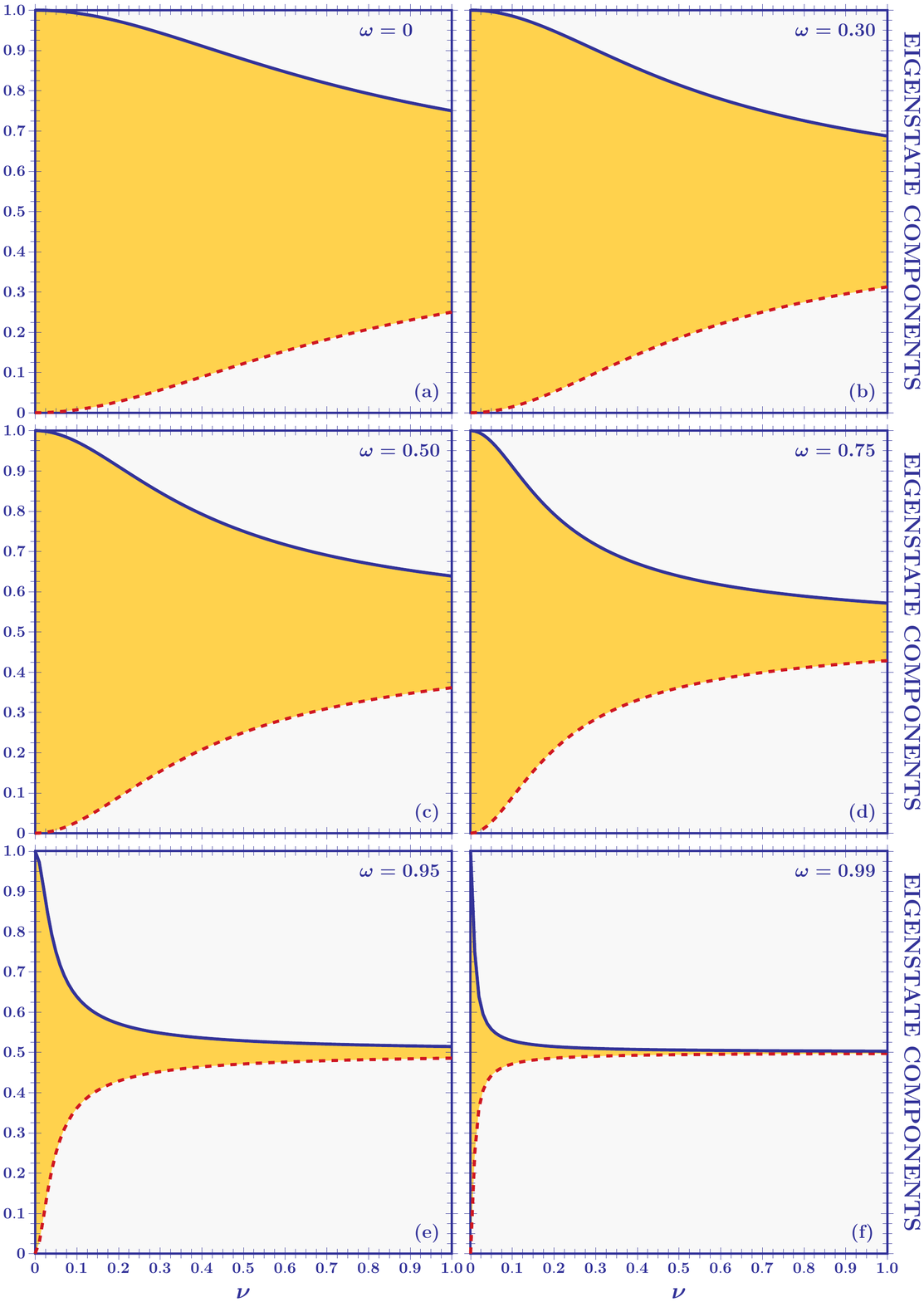}{The square of the two non-zero components of the lowest
energy eigenstate, Eq.~(\ref{eq:estate3}), as a function of $\nu$ for six values
of $\omega$. The system is composed by 3 particles.
The yellow areas emphasise the differences between the two components.\label{fig:fig3}}

\ColumnFigure{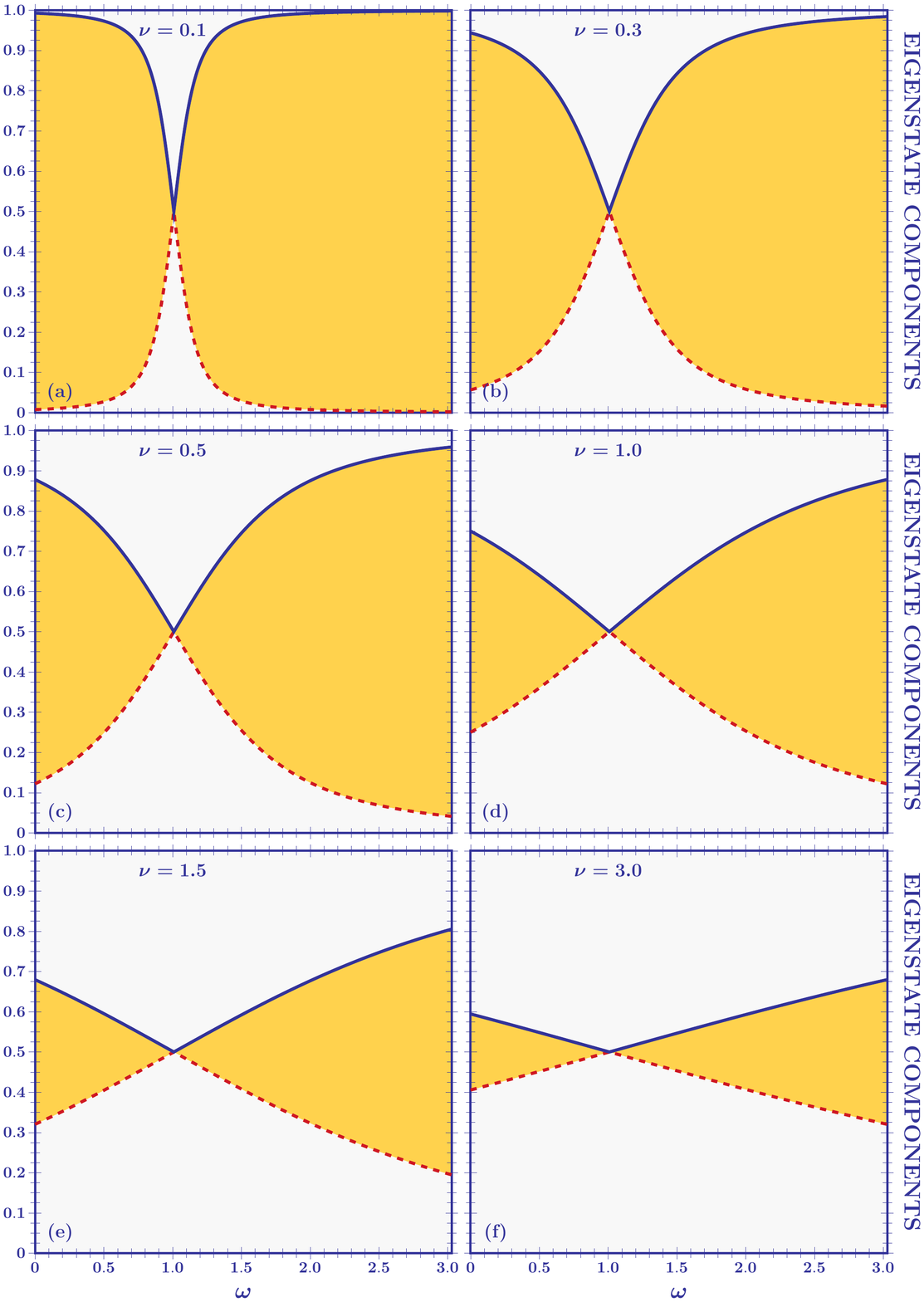}{Same as Fig.~\ref{fig:fig3} as a function of the $\omega$ strength,
for different values of $\nu$.\label{fig:fig4}}

\ColumnFigure{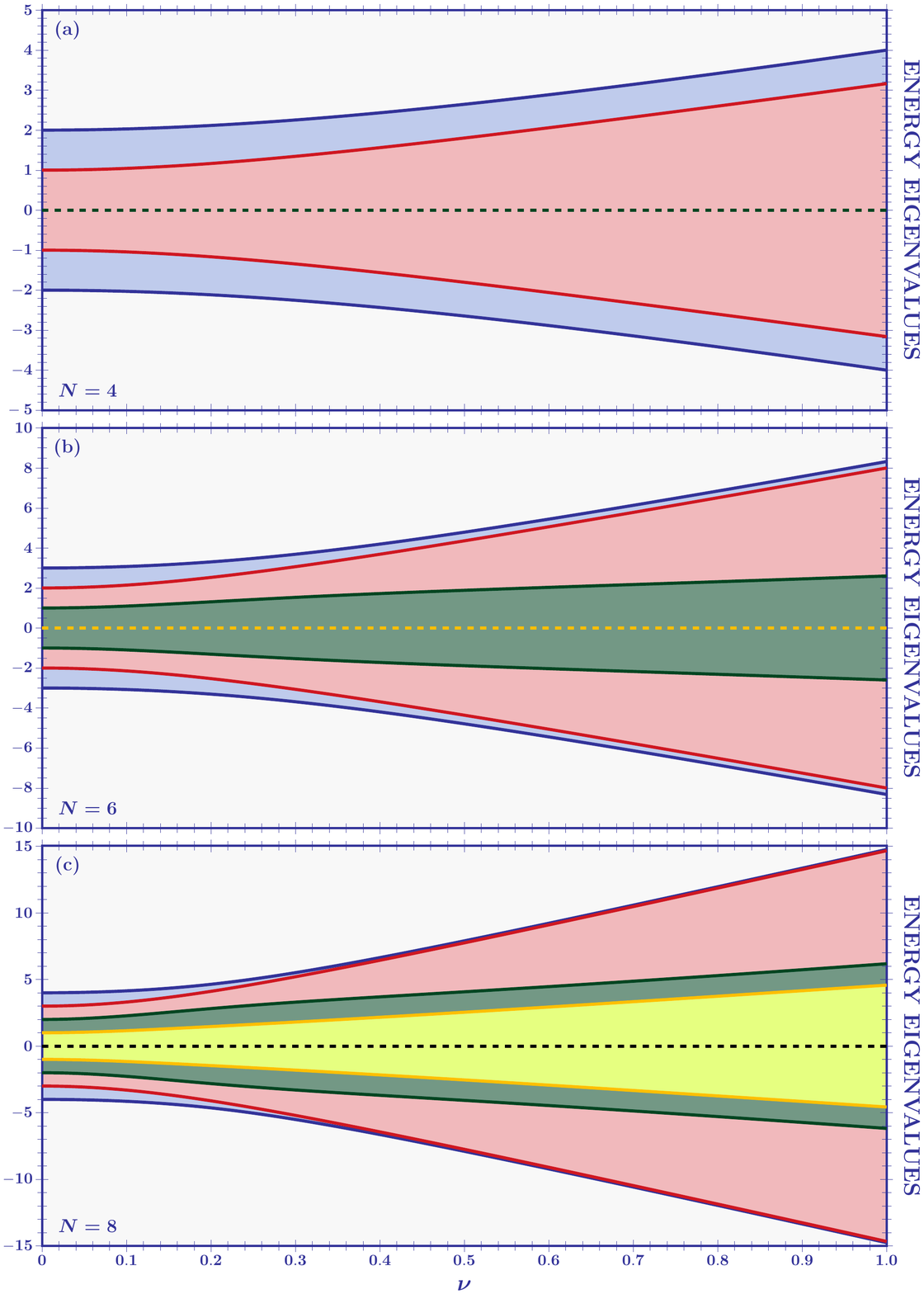}{Energy eigenvalues of the $N=4,6,8$ particle systems
as a function of $\nu$, when $\omega=0$.\label{fig:fig5}}

\ColumnFigure{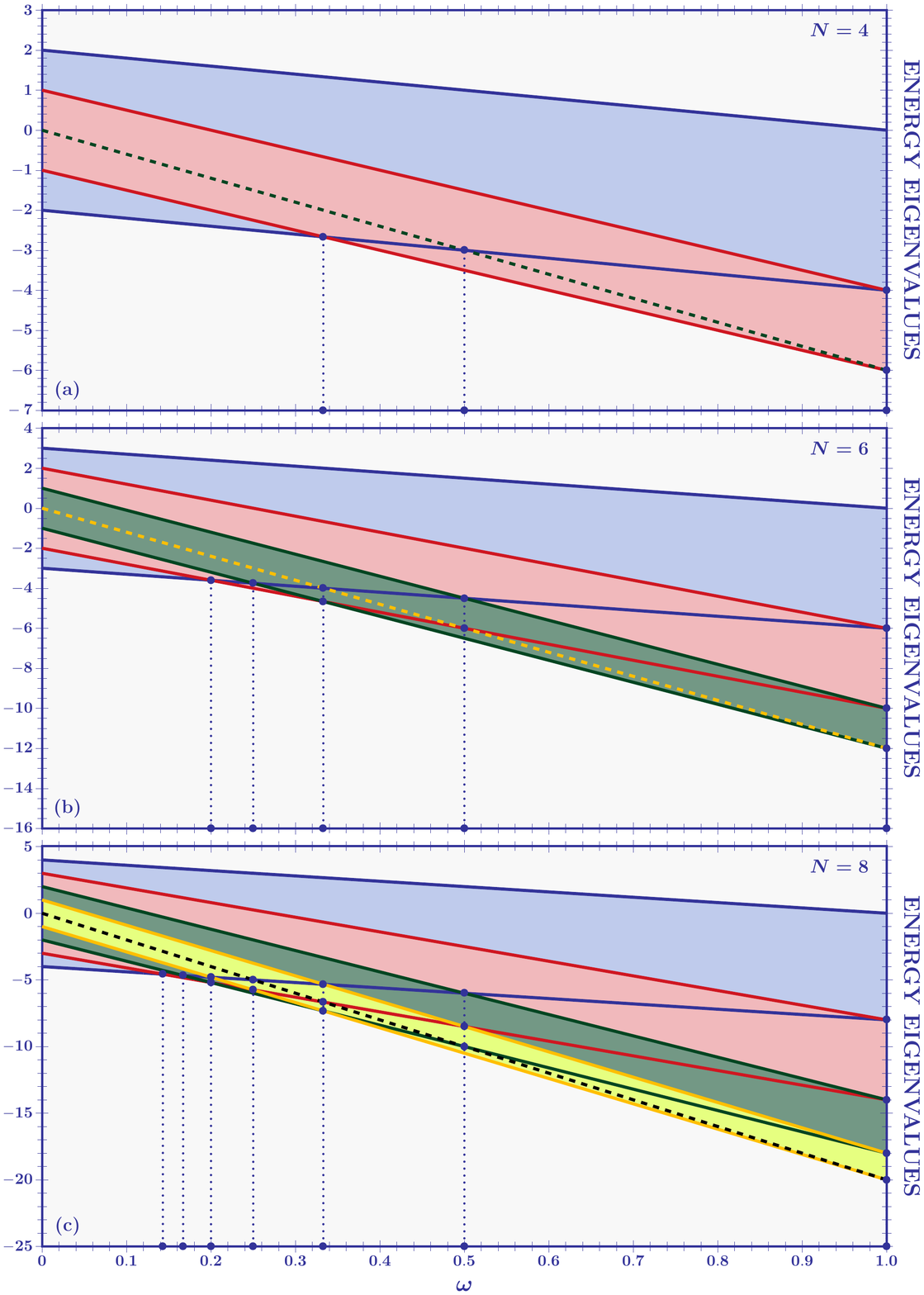}{The same as Fig.~\ref{fig:fig5}
as a function of $\omega$, when $\nu=0$. The vertical lines
indicate the degeneracy points where the energies of different
eigenstates becomes equal.\label{fig:fig6}}

\ColumnFigure{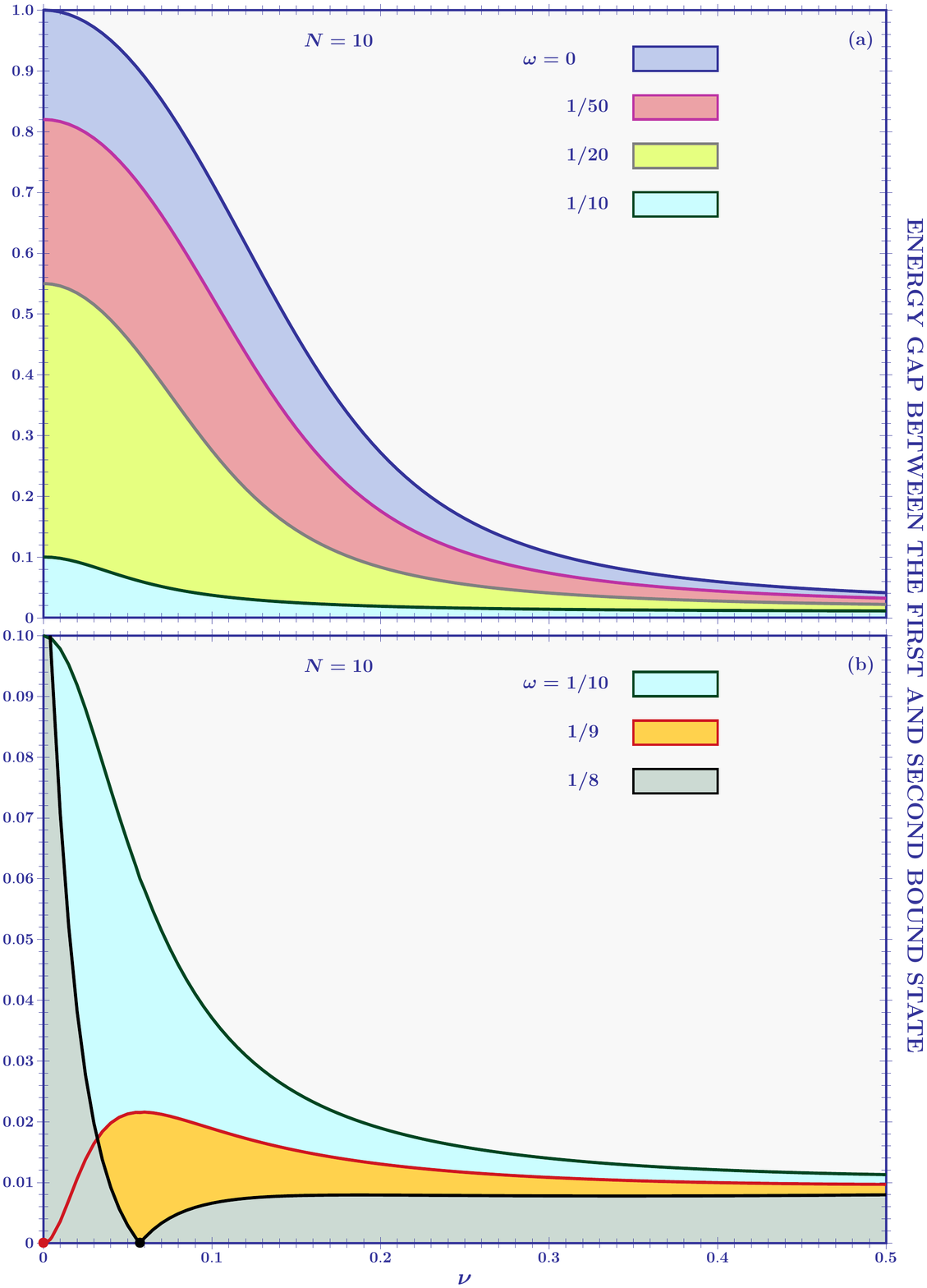}{Energy difference between the lowest two states
$\Delta_{_1.2}$ of a system of
10 particles for different values of $\omega$, as a function of $\nu$.
The different coloured areas emphasize the differences.\label{fig:fig7}}

\ColumnFigure{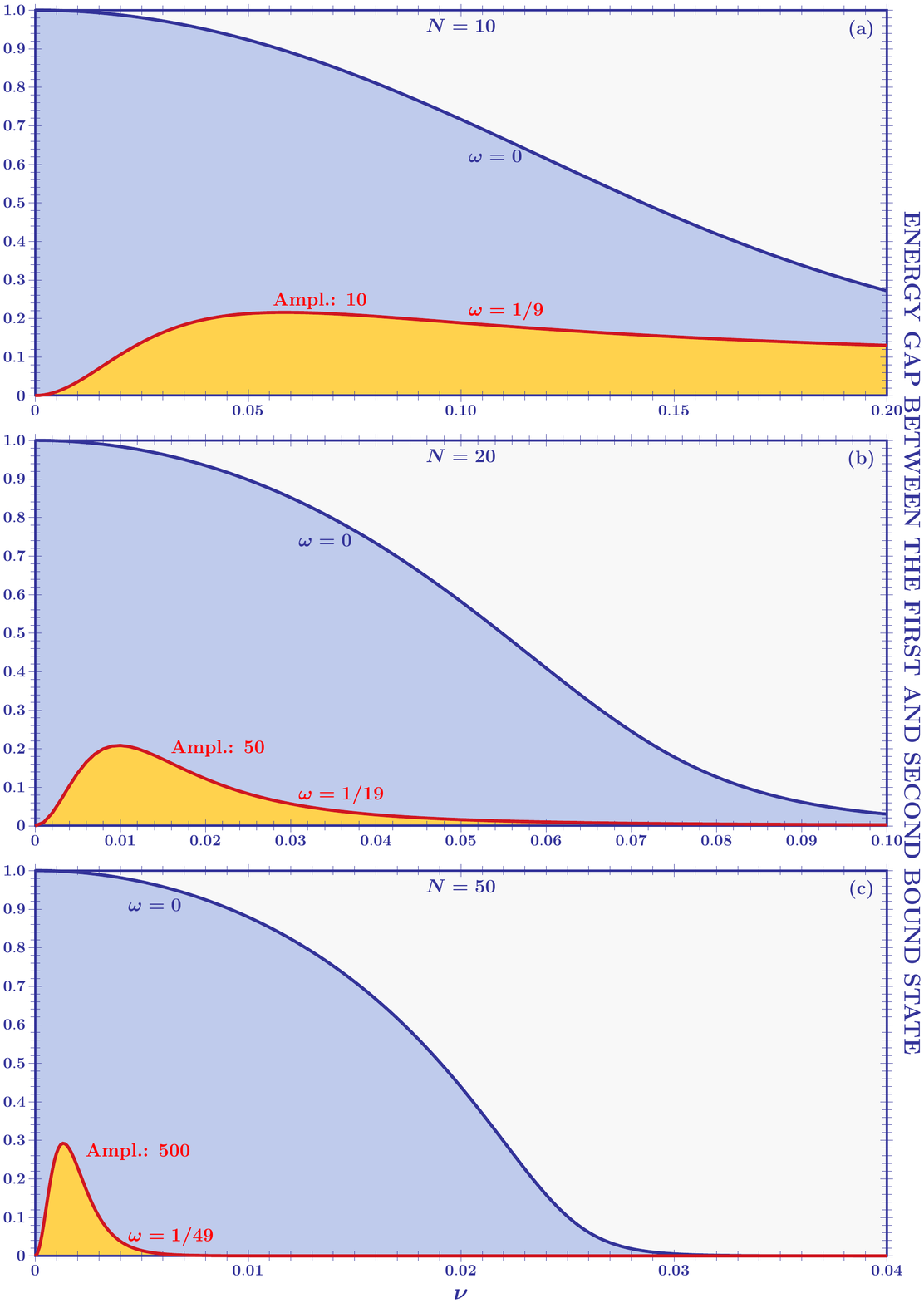}{Energy difference $\Delta_{12}^{[N]}$ between the two lowest
states of systems with $N=10,20,50$ particles.
The blue lines show results obtained with $\omega=0$ while the red
lines for  $\omega=1/(N-1)$. We point out the different $\nu$ scales
in the various panels, and the amplification factors of the red
lines.\label{fig:fig8}}

\end{document}